\documentclass[cits]{PoS}

\usepackage{amssymb, graphicx, latexsym, amsmath, bbm, color}

\def\beq{\begin{equation}}

\def\eeq{\end{equation}}

\newcommand{\ud}{\mathop{}\!\mathrm{d}}

\newcommand{\U}{\mathrm{U}}

\newcommand{\Ib}[1]{\mathrm{I}_{#1}}

\title{Fine structure of the confining string in an analytically solvable 3D model}

\ShortTitle{Fine structure of the confining string in an analytically solvable 3D
model}

\author{\speaker{Davide Vadacchino}, Michele Caselle, Roberto Pellegrini\\
        Dipartimento di Fisica Teorica, Universit\`a di Torino\\
        and Istituto Nazionale di Fisica Nucleare, sezione di Torino,\\
        Via Pietro Giuria 1, I-10125 Torino, Italy\\
        E-mail: \email{vadacchi@to.infn.it}, \email{caselle@to.infn.it}, \email{ropelleg@to.infn.it}}

\author{Marco Panero\\
       Instituto de F\'{\i}sica T\'eorica UAM/CSIC, Universidad Aut\'onoma de Madrid\\
       Calle Nicol\'as Cabrera 13-15, Cantoblanco E-28049 Madrid, Spain\\
       E-mail: \email{marco.panero@inv.uam.es}}

\abstract{In $\U(1)$ lattice gauge theory in three spacetime dimensions, confinement can be analytically shown to persist at all values of the coupling. 
Furthermore, the explicit predictions for the dependence of string tension $\sigma$ and mass gap $m_0$ on the coupling allow one to tune their ratio at will. 
These features, and the possibility of obtaining high-precision numerical results via an exact duality map to a spin model, make this theory 
an ideal laboratory to test the effective string description of confining flux tubes. In this contribution, we discuss our investigation of 
next-to-leading-order corrections to the confining potential and of the finite-temperature behavior of the flux tube width. 
Our data provide a very stringent test of the theoretical predictions for these quantities and allow to test their dependence on the $m_0/\sqrt{\sigma}$ ratio.
\vspace{4cm}

\begin{flushright}
IFT-UAM/CSIC-13-104
\end{flushright}}

\FullConference{31st International Symposium on Lattice Field Theory LATTICE 2013\\
                 July 29, 2013\\
                 Mainz, Germany}

\begin{document}

\section{Introduction}

One of the most interesting recent results on the effective string description of  Lattice Gauge Theories (LGTs) is 
the proof of universality of  the first few terms of the effective action. 
This action can be written as a low-energy expansion in the number of derivatives of the transverse degrees of freedom of the string.
The first term of this expansion is simply the 2D massless free field theory \cite{Luscher:1980fr}
\beq
S[X]=S_{cl}+\frac\sigma2\int d^2\xi\left(\partial_\alpha X\cdot\partial^\alpha X
\right) ~+\dots,
\label{gauss}
\eeq
where the classical action $S_{cl}$ describes the usual perimeter-area term,
$X$ denotes the two-dimensional bosonic 
fields $X_i(\xi_1,\xi_2)$, with  $i=1,2,\dots, D-2$,     
describing the 
transverse displacements of the string with respect the configuration 
of minimal action and $\xi_1,\xi_2$ are the coordinates on the world-sheet.
The next few terms 
beyond the free 2D bosonic theory can be parametrized as follows \cite{Luscher:2004ib}
\beq
S=S_{cl}+\frac\sigma2\int d^2\xi\left[\partial_\alpha X\cdot\partial^\alpha X+
c_2(\partial_\alpha X \cdot\partial^\alpha X)^2
+c_3(\partial_\alpha X \cdot\partial_\beta X)^2+\dots\right]\,,
\label{action}
\eeq
The coefficients $c_i$ must 
satisfy a set of constraints. These constraints were first obtained by comparing the string partition function 
in different channels (``open-closed string duality'')~\cite{Luscher:2004ib,Aharony:2009gg}.  
However it was later realized~\cite{Meyer:2006qx,Aharony:2010cx,Gliozzi:2011hj} that they are a direct consequence of
the Lorentz symmetry of the underlying Yang-Mills theory.
Indeed, even if the complete $SO(1,D-1)$  
invariance is broken by the classical configuration around which we 
expand, the effective action should still respect this symmetry through a 
non-linear realization in terms of the transverse fields $X_i$. In this way 
it was shown \cite{Aharony:2010cx} that the terms with
only first derivatives  coincide with the Nambu-Goto action to all orders in the derivative expansion.
The first allowed correction to the Nambu-Goto 
action turns out to be the the six-derivative term  \cite{Aharony:2009gg}
\beq
c_4\left(\partial_\alpha\partial_\beta X\cdot\partial^\alpha\partial^\beta X
\right)\left(\partial_\gamma X\cdot\partial^\gamma X\right)
\eeq
with arbitrary coefficient $c_4$; however this term is non-trivial only 
when $D>3$.  
The fact that the first deviations from the Nambu-Goto string are of 
high order, especially in $D=3$, explains why in early Monte Carlo 
calculations \cite{Caselle:1994df,Caselle:2005xy}  good agreement with the Nambu-Goto string was observed (for an updated review of Monte Carlo results see
\cite{Lucini:2012gg,Panero:2012qx,Lucini:2013qja}).
This agreement led to the hope to be able to identify, via numerical simulations, some non trivial ``stringy''  feature of the interquark potential. For
instance  in $D=3$ the first non-trivial deviation of the Nambu-Goto action occurs at the level
of eight-derivative terms and it was recently shown \cite{Aharony:2011gb}, 
that this term is proportional to the 
squared curvature of the induced metric on the world-sheet.

However, in the last few years, thanks to the remarkable improvement in the accuracy of simulations, systematic deviations from 
the expected universal behaviour were observed well below the eight-derivatives term mentioned above
\cite{Athenodorou:2010cs, Caselle:2007yc, Caselle:2010pf}.
These deviations were observed both in  the excited string states in (3+1)-dimensional  SU(N) LGTs~\cite{Athenodorou:2010cs} and in the ground state 
(i.e. the interquark potential) in the 3D Ising gauge 
model. In this last case deviation were observed both in the torus (interface)~\cite{Caselle:2007yc} 
and in the cylinder (Polyakov loop correlators)~\cite{Caselle:2010pf} geometries.

A possible reason for these deviations could be the presence of massive modes in the world sheet due to the coupling of the effective string to glueball states. 
If this is the case then one could try to model these massive modes and include them in the fits. The main goal of this proceeding is to test this picture in the case of
compact U(1) LGT in three dimensions. This is a perfect laboratory for this type of analysis since in this model the $m_0/\sqrt{\sigma}$ ratio is not fixed but can be
tuned as a function of the coupling. Thus  by tuning $\beta$ we can compare the ability of the effective string to fit the interquark potential when the  
$m_0/\sqrt{\sigma}$ takes a value comparable with those of the Ising or SU(N) LGTs (i.e. $m_0/\sqrt{\sigma}\sim 3$) and for a much lower value   $m_0/\sqrt{\sigma}\sim 1$.
As we shall see for $m_0/\sqrt{\sigma}\sim 1$ the effective string predictions strongly disagree with the numerical data.

\section{The $\U(1)$ model in $2+1$ dimensions}

The $\U(1)$ lattice gauge model in $D=2+1$ 
can be  solved  in the semiclassical
approximation~\cite{Polyakov:1976fu, Gopfert:1981er}. It can be shown that the model is always confining 
and that in the $\beta\gg 1$ limit it flows toward a theory of free massive scalars. 
In this limit the mass of the lightest glueball and 
the string tension should behave as
\begin{equation}
m_0 = c_0\sqrt{8 \pi^2 \beta}e^{-\pi^2\beta v(0)},\;\;\; \sigma \geq
\frac{c_{\sigma}}{\sqrt{2\pi^2\beta}} e^{-\pi^2\beta v(0)},
\end{equation}
where in the semiclassical approximation $c_0=1$ and $c_{\sigma}=8$.
Previous numerical studies \cite{Loan:2002ej} (which we confirm with our simulations) 
showed that the string tension saturates the above bound and that
both constants are affected by the semiclassical approximation and change their values in the continuum limit. 
Despite this difference, both $m_0$ and $\sigma$ remain strictly positive in the whole range of values of $\beta$, the model is
thus always confining. The point in using such lattice model at finite
spacing is that, while in general for confining gauge theories the $m_0/\sqrt{\sigma}$ ratio is fixed, in this model
we have:
\begin{equation}
\frac{m_0}{\sqrt{\sigma}} = \frac{c_0}{\sqrt{c_\sigma}} 2\pi
(2\pi\beta)^{3/4} e^{-\pi^2 v(0)\beta/2}
\end{equation}
so that we can tune the ratio to any chosen value  by tuning $\beta$.

Since the model is invariant under an Abelian gauge symmetry, one can
easily perform a duality transformation and obtain a simple spin model
with global $\mathbb{Z}$ symmetry (note that, in $D=3+1$ dimensions, the
same transformation leads to a model with local $\mathbb{Z}$
symmetry~\cite{Zach:1997yz, Panero:2005iu, Panero:2004zq,
Cobanera:2011wn, Mercado:2013ola}). The dual formulation has many advantages over the
original one. First of all, from the computational point of view, the
model is much easier and faster to simulate, since we deal with a spin
model. Moreover, the presence of a static $Q\bar{Q}$ pair at a distance
$R$ in the form of Polyakov loops can be easily included in the partition
function that becomes, in that case
\begin{equation}
Z_R = \sum_{\left\{ \,^\star l=-\infty\right\}}^{\infty} \prod_{^\star
c_1} \Ib{|\ud^\star l+^\star n|}(\beta)
\end{equation}
where $n$ is a 2-form which must be non-vanishing on an arbitrary surface
bounded by the two loops; without loss of generality, we choose the one of
minimal area.

\section{Results}

We performed our simulations on the dual model, using a single site
metropolis algorithm, at three different 
values of the coupling, on $(64a)^2\times L$ lattices, with $L$ the lattice size
in the Euclidean time direction. We chose two values $L=16a$ and $L=64a$ and evaluated the interquark potential for all values of
the interquark distance $R$ in the range $8a \leq R\leq 32a$ so as to have a wide range of values of $R$, 
such that $2R<L$ and $2R>L$ for $L=64a$ and $L=16a$ respectively. As we shall see below in these two limits the 
effective string predictions
for the interquark potential simplify and fitting the simulation results becomes much simpler.
The values of the coupling were chosen so that the $m_0/\sqrt{\sigma}$ ratio ranged from $\sim2.5$ for $\beta=1.7$ to $\sim 0.7$ for
$\beta=2.75$. Details on the simulation setting are reported in tab.~\ref{tab:info}.

The simulations were performed on the dual model using the snake
algorithm~\cite{deForcrand:2000fi} with a hierarchical update
algorithm~\cite{Caselle:2002ah} to decorrelate data.

Following \cite{{Caselle:2005xy},Caselle:2002ah} we
computed the quantity
\begin{equation}
Q(R) = - \frac{a}{L} \log{\left[ \frac{G(R+a)}{G(R)} \right]} = a\left[ V(R)-V(R+a) \right]
\end{equation}
in which the perimeter  and costant terms of the
interquark potential cancel. 
Setting in eq. (\ref{action}) the values $c_2=\frac18$ and $c_3=-\frac14$ prescribed by Lorentz invariance, we find
at next-to-leading order \cite{Caselle:2005xy},
\begin{equation}\label{eq:qzerofinite}
Q(R) =
\begin{cases}
\sigma a^2 + \frac{\pi a^2}{24 R (R+a)} + \frac{\pi^2 a^2}{1152 \sigma}\frac{2 R^2 +
2aR + a^2}{R^3 (R+a)^3} + \cdots\, & 2R\ll L\\
\sigma a^2 \left( 1 - \frac{\pi}{6 L^2 \sigma} - \frac{\pi^2}{72\sigma^2 L^4}+ \cdots
\right)+ \frac{a^2}{2LR} -\frac{a^2}{8\sigma L^2 R^2} + \cdots\, & 2R\gg L\\
\end{cases}
\end{equation}

\begin{figure}[-t]
\centerline{\includegraphics[width=\textwidth]{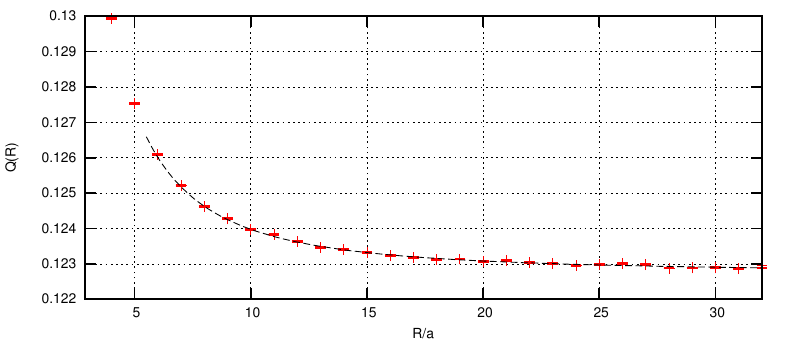}}
\centerline{\includegraphics[width=\textwidth]{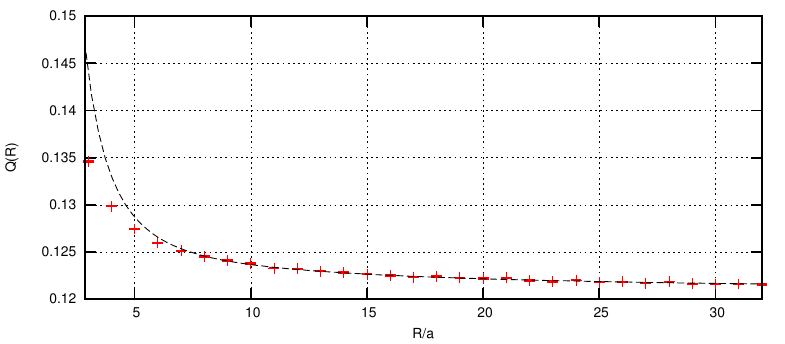}}
\caption{Results from simulations at $\beta = 1/(a e^2)= 1.7$. For this
coupling, $m_0/\sqrt{\sigma} \simeq 2.5$ and $1/(a\sqrt{\sigma}) \simeq 3$. The
values of $L$ and $R$ are chosen so that $R<L/2$ for the plot in the top panel,
$R>L/2$ for the one  at the bottom. The curves correspond to the best fit results 
obtained with eqs.~(\protect\ref{eq:qzerofinite}) setting $R_{min}=12a$.
\label{fig:effstr1}}
\end{figure}

\begin{figure}[-t]
\centerline{\includegraphics[width=\textwidth]{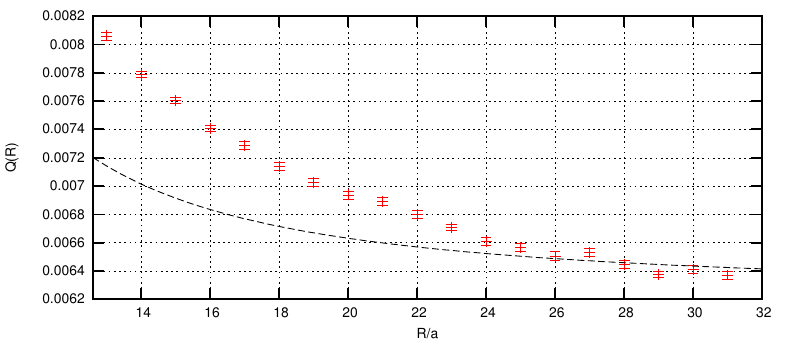}}
\centerline{\includegraphics[width=\textwidth]{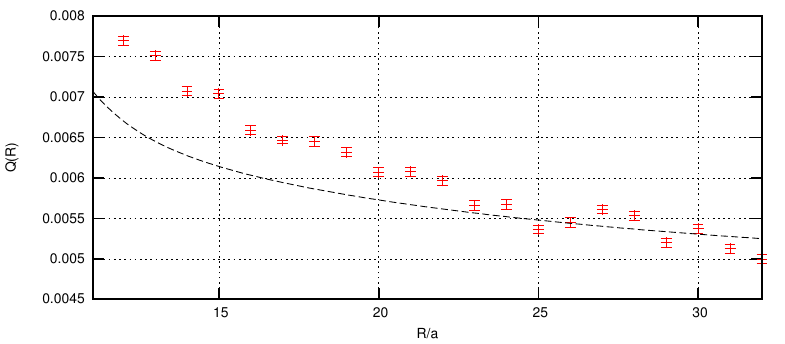}}
\caption{Same as in fig.~\protect\ref{fig:effstr1}, but for  $\beta = 2.75$ (for which $m_0/\sqrt{\sigma} \simeq 0.7$ and $1/(a\sqrt{\sigma}) \simeq 12$) and $R_{min}=24a$.
\label{fig:effstr2}}
\end{figure}

We fitted the data obtained in the simulations with eqs.~\ref{eq:qzerofinite} keeping $\sigma a^2$ as the only free parameter 
and used the reduced $\chi^2$ to estimate the quality of the
various fits. We performed two types of fits. In the first we kept for all values of $\beta$ 
exactly the same number of 
degrees of freedom in the fit, fitting for both $L=16a$ and $L=64a$ all the data in the range $12a \leq R \leq 32a$. In the second 
we fitted the data in the range $R_{min}\leq R \leq 32a$ choosing different thresholds $R_{min}$ so as to keep the adimensional ratio
$\sqrt{\sigma} R_{min}\sim 2$ for each value of $\beta$. This second choice decreases the number of degrees of 
freedom for the larger values of $\beta$ but eliminates the
bias due to the difference in the value of $1/(a\sqrt{\sigma})$ in the three samples.    
Results are reported in tabs.~\ref{tab:fit1} and ~\ref{tab:fit2}, respectively. We also plot in fig.~\ref{fig:effstr1} and fig.~\ref{fig:effstr2} the result of the
fit in the two limiting cases $\beta=1.7$ and $\beta=2.75$.

\begin{table}[ht]
\centering
\begin{tabular}{|c|c|c|}
\hline
$\beta$ & ${m_0}/{\sqrt{\sigma}}$ & $\sigma\,a^2$\\
\hline
$1.7$ & $2.5(1)$   & $0.1227(1)$ \\
\hline
$2.2$ & $1.6(2)$  & $0.0273(1)$ \\
\hline
$2.75$ & $0.7(1)$ & $0.0070(1)$ \\
\hline
\end{tabular}
\caption{Values of ${m_0}/{\sqrt{\sigma}}$  
and  $\sigma a^2$ for the three values of $\beta$. While the dependence on $\beta$ follows the semiclassical prediction, the constant in front of $\sigma$ and $m_0$ differ from
the semiclassical ones }
\label{tab:info}
\end{table}

\begin{table}[ht]
\centering
\begin{tabular}{|c||c|c|c|c|}
\hline
$\beta$ & $L=64\,a$  & $L=16\,a$ &  $R_{min}/a$ & d.o.f. \\
\hline
$1.7$ & $1.1$  &  $1.2$ & $12$  & $20$  \\
\hline
$2.2$ & $2.2$  &  $9.9$ & $12$  & $20$ \\
\hline
$2.75$ & $47.4$  & $178.6$ & $12$ & $20$ \\
\hline
\end{tabular}
\caption{Values of $\chi^2_r$ obtained by comparing the values $Q(R)$
obtained in lattice simulations with the prediction of effective string
theory at NLO, for $R\geq R_{min}$.}
\label{tab:fit1}
\end{table}

\begin{table}[ht]
\centering
\begin{tabular}{|c||c|c|c|c|}
\hline
$\beta$ & $L=64\,a$  & $L=16\,a$ &  $R_{min}/a$ & d.o.f.  \\
\hline
$1.7$ & $1.1$  &  $1.2$ & $8$  & $24$   \\
\hline
$2.2$ & $1.8$  &  $2.2$ & $16$  & $16$  \\
\hline
$2.75$ & $9.4$  & $5.1$ & $24$ & $8$  \\
\hline
\end{tabular}
\caption{Same as tab.~\protect\ref{tab:fit1} but for different values of $R_{min}$ for the different values of $\beta$.}
\label{tab:fit2}
\end{table}

Looking at tabs.~\ref{tab:fit1} and \ref{tab:fit2}  and at figs.~\ref{fig:effstr1} and ~\ref{fig:effstr2} 
we see that for $\beta=1.7$ (for which $m_0/\sqrt{\sigma}\sim 2.5$) the effective string describes very precisely the data 
while for $\beta=2.75$ (for which $m_0/\sqrt{\sigma}\sim 0.7$) there is clear disagreement.
In this second case even the first order effective string correction, the L\"uscher term, is excluded by the data. This behaviour is in
striking disagreement with the results of almost all the numerical experiments performed 
in the last few years in other lattice gauge theories where the
L\"uscher term was always found to agree with the data for large enough values of $R$.

These results show that the ability of the effective string to describe the interquark potential is strongly related to the ratio 
$m_0/\sqrt{\sigma}$, and worsens as the lowest glueball mass approaches
$\sqrt{\sigma}$. This preliminary test suggests that indeed  the emission of glueballs can masks
the stringy behaviour of the system and could be the ultimate reason 
of the discrepancies observed in refs.~\cite{Athenodorou:2010cs, Caselle:2007yc, Caselle:2010pf}. 
Similarly to other 3D models~\cite{Athenodorou:2011rx}, the 3D U(1) model could be a very powerful tool 
to study the interplay between effective strings and glueballs.

\paragraph{Acknowledgments}

This work is supported by the Spanish MINECO's ``Centro de Excelencia
Severo Ochoa'' programme under grant SEV-2012-0249.


\clearpage

\begin{thebibliography}{99}



\bibitem{Luscher:1980fr}
M.~L\"uscher, K.~Symanzik and P.~Weisz, 
{\it Nucl.\ Phys.} {\bf B173}, 365 (1980).

\bibitem{Luscher:2004ib}
M.~L\"uscher and P.~Weisz,
{\it JHEP} {\bf 07}, 014 (2004),
{\tt [arXiv:hep-th/0406205]}.

\bibitem{Aharony:2009gg}
O.~Aharony and E.~Karzbrun, 
{\it JHEP} {\bf 06}, 012 (2009),
{\tt [arXiv:0903.1927 [hep-th]]}.

\bibitem{Meyer:2006qx}
H.~B. Meyer, 
{\it JHEP} {\bf 05}, 066 (2006),
{\tt [arXiv:hep-th/0602281]}.

\bibitem{Aharony:2010cx}
O.~Aharony and M.~Field, 
{\it JHEP} {\bf 1101}, 065 (2011), 
{\tt [arXiv:1008.2636 [hep-th]]}.

\bibitem{Gliozzi:2011hj}
F.~Gliozzi, 
{\it Phys. \ Rev.} {\bf D84}, 027702 (2011),
{\tt [arXiv:1103.5377 [hep-th]]}.

\bibitem{Caselle:1994df}
M.~Caselle {\em et al.},
{\it Nucl. \ Phys.} {\bf B432}, 590 (1994), 
{\tt [arXiv:hep-lat/9407002]}.

\bibitem{Caselle:2005xy}
M.~Caselle, M.~Hasenbusch and M.~Panero, 
{\it JHEP} {\bf 03}, 026 (2005),
{\tt [arXiv:hep-lat/0501027]}.

\bibitem{Lucini:2012gg}
  B.~Lucini and M.~Panero,
  {\it Phys.\ Rept.\ } {\bf 526} (2013) 93,
  {\tt [arXiv:1210.4997 [hep-th]]}.

\bibitem{Panero:2012qx}
  M.~Panero,
  \pos{PoS(LATTICE 2012)010},
  {\tt [arXiv:1210.5510 [hep-lat]]}.

\bibitem{Lucini:2013qja}
  B.~Lucini and M.~Panero,
  {\tt arXiv:1309.3638 [hep-th]}.

\bibitem{Aharony:2011gb}
O.~Aharony and M.~Dodelson,
{\it JHEP} {\bf 1202} (2012) 008,
{\tt [arXiv:1111.5758 [hep-th]]}.

\bibitem{Athenodorou:2010cs}
A.~Athenodorou, B.~Bringoltz and M.~Teper, 
{\it JHEP} {\bf 02}, 030 (2011),
{\tt [arXiv:1007.4720 [hep-lat]]}.

\bibitem{Caselle:2007yc}
  M.~Caselle, M.~Hasenbusch and M.~Panero,
  {\it JHEP} {\bf 0709} (2007) 117,
  {\tt [arXiv:0707.0055 [hep-lat]]}.

\bibitem{Caselle:2010pf}
  M.~Caselle and M.~Zago,
  {\it Eur.\ Phys.\ J.}\  {\bf C71 } (2011)  1658,
  {\tt [arXiv:1012.1254 [hep-lat]]}.

\bibitem{Polyakov:1976fu}
  A.~M.~Polyakov,
  {\it Nucl.\ Phys.\ } {\bf B120} (1977) 429.

\bibitem{Gopfert:1981er}
  M.~G\"opfert and G.~Mack,
  {\it Commun.\ Math.\ Phys.\ }  {\bf 82} (1981) 545.

\bibitem{Loan:2002ej}
  M.~Loan {\em et al.},
  {\it Phys.\ Rev.\ } {\bf D68} (2003) 034504,
  {\tt [arXiv:hep-lat/0209159]}.

\bibitem{Zach:1997yz}
  M.~Zach, M.~Faber and P.~Skala,
  {\it Phys.\ Rev.\ }{\bf D57} (1998) 123,
  {\tt [arXiv:hep-lat/9705019]}.

\bibitem{Panero:2005iu}
  M.~Panero,
  {\it JHEP} {\bf 0505} (2005) 066,
  {\tt [arXiv:hep-lat/0503024]}.

\bibitem{Panero:2004zq}
  M.~Panero,
  {\it Nucl.\ Phys.\ Proc.\ Suppl.\ }  {\bf 140} (2005) 665,
  {\tt [arXiv:hep-lat/0408002]}.

\bibitem{Cobanera:2011wn}
  E.~Cobanera, G.~Ortiz and Z.~Nussinov,
  {\it Adv.\ Phys.\ } {\bf 60} (2011) 679,
  {\tt [arXiv:1103.2776 [cond-mat.stat-mech]]}.

\bibitem{Mercado:2013ola}
  Y.~D.~Mercado, C.~Gattringer and A.~Schmidt,
  {\it Phys.\ Rev.\ Lett.}\  {\bf 111} (2013) 141601,
  {\tt [arXiv:1307.6120 [hep-lat]]}.

\bibitem{deForcrand:2000fi}
  Ph.~de Forcrand, M.~D'Elia and M.~Pepe,
  {\it Phys.\ Rev.\ Lett.\ } {\bf 86} (2001) 1438,
  {\tt [arXiv:hep-lat/0007034]}.

\bibitem{Caselle:2002ah}
  M.~Caselle, M.~Hasenbusch and M.~Panero,
  {\it JHEP} {\bf 0301} (2003) 057,
  {\tt [arXiv:hep-lat/0211012]}.

\bibitem{Athenodorou:2011rx}
  A.~Athenodorou, B.~Bringoltz and M.~Teper,
  {\it JHEP} {\bf 1105} (2011) 042,
  {\tt [arXiv:1103.5854 [hep-lat]]}.

\end{thebibliography}
\end{document}